\documentclass[traditabstract]{aa} 

\usepackage{graphicx}
\usepackage{txfonts,natbib,url}
\bibpunct{(}{)}{;}{a}{}{,}    
\begin{document}
  \title{Size distribution of supernova remnants and the interstellar medium: the case of M33}
   \author{Abdul I. Asvarov}
   \offprints{A.I.Asvarov}
   \institute{Institute of Physics, Azerbaijan National Academy of Sciences, BAKU, Azerbaijan\\
              \email{asvarov@physics.ab.az}}          
   \date{Received month day, year; accepted month day, year}   
   \authorrunning{Asvarov A.I.}
   \titlerunning{Size distribution of SNRs}
\abstract { 
The size distribution of supernova remnants (SNRs) can help to clarify the various aspects of their evolution and interaction with the interstellar medium (ISM). Since the observed samples of SNRs are a collection of objects with very different ages and origin that evolve in different conditions of the ISM, statistical Monte Carlo methods can be used to model their statistical distributions.  Based on very general assumptions on the evolution, we have modeled samples of SNRs at various initial and environmental conditions, which were then compared with observed collections of SNRs. In the evolution of SNRs the pressure of the ISM is taken into account, which determines their maximum sizes and lifetimes.  When comparing the modeled and observed distributions, it is very important to have homogeneous observational data free from selection effects.  We found that a recently published collection of SNRs in M33 
satisfies this requirement if we select the  X-ray SNRs with hardness ratios in a limited range of values. An excellent agreement between distributions of this subset of SNRs and the subset of modeled SNRs was reached for a volume filling-factor of the warm phase of the ISM (partly ionized gas with $n_{\rm H}\sim 0.2-0.5~ \rm {cm}^{-3}; T \sim 8000-10000~K $)
in M33 of $\sim\ 90\%$. The statistical distributions constructed in this way, which reproduce practically all the statistical properties of observed SNRs, allowed us to obtain one of the important parameters of M33: the birthrate is one SNR every $ {140} - {150}$ yr, and the total number of SNRs with a  shock Mach number $M_{s} \geq 2$ is larger than $\sim 1000$.}

   \maketitle

\section{Introduction}

Understanding the size distribution of supernova remnants (SNRs) can be useful to study the physics and astrophysics of these objects themselves as well as the interstellar medium (ISM) of the galaxy where their evolution takes place. 
The form of the size distribution of SNRs is determined by the properties of their exploding parent stars, the mechanism of supernova (SN) events,  and  the properties of the interstellar medium where SNRs  are evolving. The majority of the observed SNRs consists of older, evolved SNRs, but they still retain information on the energy released by their exploding progenitors. 
 Although supernovae are divided into two large groups, SNI and SNII, which are quite different in the properties of parent stars and the mechanisms of explosion, they deposit a similar amount ($\sim10^{51}$ erg) of kinetic energy into the surrounding medium, therefore we did not distinguish between types of SNe in this study.
 Other factors affecting  the size distribution of SNRs are the physical properties of the interstellar medium, mainly the density and, in the final stages of the evolution, the total pressure  of the ISM. The pressure in the ISM is the main factor that determines the final fate of the SNR. 

\citet{Mathewson1983} were among the first to report on the size distribution of SNRs in the Magellanic Clouds, and from the approximately linear cumulative size distribution the authors have concluded that a majority of SNRs in the Magellanic Clouds are in a ``free expansion'' phase of evolution, with shocks propagating at a constant velocity into a tenuous ambient medium. This conclusion was in apparent contradiction with the observational evidence that a large portion of observed SNRs in the Magellanic Clouds are in an adiabatic phase. The effect of the various selection effects and the nonuniformity of the ISM density distribution  on the size distributions are considered in a number of works \citep[e.g.,][]{Hug84,Berk1986,Berk1987,FF84,Bad2010}.

Obviously, the detected and cataloged SNRs constitute only a small fraction of the real number of objects. This fraction  depends on various factors -   the waveband of observations, the characteristics of the instrument with which the object is observed, on the site where the remnants are located, etc. Since the observed sample of SNRs consists of objects with different and  uncorrelated parameters,  and because they evolve in environments with different properties, the use of Monte Carlo methods is justified and can be very effective in  statistical studies of SNRs. Using the real bands of variation of the physical parameters that characterize the remnant and the ISM, we generated a set of SNRs that can be compared with the homogenous subset of observed SNRs with known parameters. The realization of this scheme became possible thanks to the recently published results by  \citet{L10} (hereafter L10), which present very rich and high-quality Chandra  X-ray observational data on the SNRs in the nearby galaxy M33 from the deepest Chandra ACIS survey of M33 (ChASeM33).

The main aim of the present study is to improve our understanding of the problem of the evolution of SNRs in real conditions of the ISM. This complex problem consists of a number of aspects - the law of expansion, the evolution of emission in various bands of the electromagnetic spectrum, the interrelation of the observational characteristics of SNRs with the properties of the ISM where the objects are located, etc.  

The paper is organized as follows: in Sect. 2 we consider the evolution of the SNR, where we derive the equations describing the dependence of the shock Mach number on the radius of the remnant with and without cooling.  In Sect. 3 we present the observational X-ray data of SNRs in M33. In this section we also discuss the thermal X-ray emission from evolved SNRs. 
From the catalog of SNRs in M33, we have identified an ensemble of objects that can be considered as a statistically complete set, which is then used in Section 4 for comparison with the Monte Carlo modeled SNRs. In  Section 4 we also present an analysis of the results.
  The final section contains our main conclusions.

\section{Expansion of SNRs}

The problem of expansion of the supernova blast wave in general is well understood and described in number of detailed studies \citep[e.g.,][]{Zel66,Bis95,OM88}. In the course of evolution with time, the SNR evolves through several phases. The first, free expansion phase lasts up to the moment when the swept-up mass equals the ejecta mass. Since the duration of this phase is relatively short, fewer than $\sim 10^{3}$ years, this period of life of the remnant plays only a minor role in the overall statistics of SNRs. By the time the swept-up mass equals the ejecta mass a smooth transition from this stage to the adiabatic Sedov-Taylor phase begins \citep{TM99}.  This phase is believed to play very important role in the life of the remnant because at this stage SNR intensively emits practically in all wavelengths of the electromagnetic spectrum from radio to gamma. The structure of the remnant is described by the exact self-similar Sedov' solution. The transition to this phase is smooth and asymptotic \citep{cioffi88}; depending on the real conditions, the duration and applicability of this solution to real SNR may vary within broad limits. In all cases, the sudden liberation of a large amount of energy in a small volume would result in generation of strong shock wave with the structure asymptotically approaching to the self-similar Sedov solution \citep{LL,Zel66}. The radius of the spherical blast wave that evolves in the homogeneous and stationary interstellar medium changes according to the expansion law \citep{s77}

\begin{equation}
	 R_{\rm s} = {\left( {\frac{{{\xi }\,{E_0}}}{{{\rho _0}}}} \right)^{1/5}}{t^{2/5}},
\label{eq1}
\end{equation}
where ${R_s}$ is the radius of the shock front, ${E_0}$ is total (thermal plus kinetic) energy of  the blast wave, ${\rho _0}$ is the mass density in the ISM where the shock front evolves, ${\xi }$ is the dimensionless numerical constant, which depends on the adiabatic index of the plasma $\gamma$: $\xi = 2.026$ for $\gamma = 5/3$.

To ensure a smooth transition from free expansion to the expansion law (1) we used the condition of constancy of the kinetic energy of the shell with an ejected mass of $M_{\rm{ej}}$: 
\begin{displaymath}
\left( {{M_{\rm{ej}}} + \frac{4}{3}\pi {\rho _0}R_s^3} \right) \cdot {\left( {\frac{{d{R_s}}}{{dt}}} \right)^2} = 2{E_{0k}},
\end{displaymath}
 which gives for the shock wave velocity
\begin{equation}
\upsilon_{\rm s} = \frac{{{\upsilon_{\rm 0s}}}}{{\sqrt {1 + {{({R_s}/{R_0})}^3}} }}, 
\label{eq2}
\end{equation}
where ${R_0} = {\left( {3{M_{{\rm{ej}}}}/4\pi {\rho _0}} \right)^{1/3}}$,  and  ${\upsilon _{\rm{0s}}} = \sqrt {2{E_{{\rm{0k}}}}/{M_{{\rm{ej}}}}}$, from which it follows that when $R_s<<R_0$, $\upsilon_{\rm{s}}=\upsilon_{\rm{0s}}=const$ (free expansion) and at $R_{\rm{s}} >> R_0$,   $\upsilon_{\rm{s}}=\upsilon_{\rm{0s}}\cdot (R_{\rm{s}}/R_0)^{-3/2}$ in accordance with the Sedov expansion law (Eq.~\ref{eq1}).
 
In Eq.~(\ref{eq2}) the evolution of the SNR  is described with the help of the dependence of the forward shock velocity on the radius  instead of the standard radius - time (age) relation as in Eq.~(\ref{eq1}).  The evolution of the remnant can also be described  with the help of other dependences, for instance, velocity\ - age (time), velocity-radius relationships. The Mach number of the SNR shock wave can be used as another parameter to describe the evolution of the supernova remnant.  This quantity allows us to include the parameters of the ISM into the expansion equation. 
We determine the Mach number as the ratio of the shock velocity ${\upsilon _{\rm{s}}} = d{R_{\rm{s}}}/dt$   to the maximal speed of propagation of small disturbances in the magnetized ISM, (magneto-) sound speed, $c _{\rm ms0}$:    
\begin{equation}
M_{\rm s} =\frac{\upsilon_{\rm s}}{c_{\rm ms0}}. 
\label{eqMN}
\end{equation} The (magneto-) sound speed in the ISM is determined as ${c_{\rm ms0}} = \sqrt {\gamma {P_0}/{\rho _0}} $, where $P_0$ is the total pressure (the sum of thermal and nonthermal particles and magnetic field pressures) in the ISM. Note that including the relativistic gas component and the magnetic fields leads to a softening the equation of state of the plasma, and the adiabatic index $\gamma$ is expected to be in the range 4/3 - 5/3. 
	
	The applicability of the Sedov self-similar solution to real SNR is determined by two main conditions: 1) the pressure of the ambient medium is negligible and   2)   the condition of adiabaticity of the matter inside the remnant holds. Violation of any of these conditions leads to a violation of the applicability of the Sedov solution.  In very tenuous environments the adiabaticity is retained up to the very large diameters at which the pressure of the interstellar medium becomes important. It is important to note that the law of motion (Eq.~\ref{eq1}) gives relatively good results up to values of the shock Mach numbers of $2$, although the internal structure of the shell begins to depart from the Sedov self-similar solution much earlier, when $M_{\rm s}\leq 10$ \citep{s77,CA82}.  
In relatively dense medium the decrease of the post-shock plasma temperature with time increases the radiative losses of the matter in the shell of the remnant. According to \citet {cox72},  at some time during the evolution  (moment $t_{\rm sg}$, sag time),  radiative cooling begins to affect the temperature distribution downstream and initiates the deviation from the self-similar nature of the flow. 
For an SNR evolving in a homogeneous ISM, \citet{cioffi88} found that the time taken to cool an element of mass to $0$ temperature, that is, to form the  shell, is  

\begin{equation}
{t_{{\rm{sf}}}} = 3.61 \times {10^4}\zeta _m^{ - 5/14}E_{51}^{3/14}n_0^{ - 4/7}\,{\rm{years}},	
	\label{eq3}
\end{equation}
where ${E_{51}}$ is the supernova energy in units of ${10^{51}}$  ergs,  ${n_{0}}$ is the number density of the gas in the ISM, ${\zeta _m}$ is the metallicity factor (=1 for solar abundances). After about this time the shock can be described by the radiative pressure-driven snowplow (PDS) model. In a number of papers the problem of transition from the adiabatic to the radiative phase of evolution is discussed. Practically the same expression as Eq.~(\ref{eq3}) for the transition time is derived in \citet{CA82}, although they used slightly different methods. In \citet{franco} the same time is derived with the same functional form, but with a slightly different value of the coefficient. 
 A more detailed analysis of the problem of transition of the SNR from the adiabatic phase to the radiative phase is given by \citet{petruk06}.

Assuming that the remnant evolves according to the Sedov solution up to the very moment when the shell formation finishes, $t_{\rm{sf}}$,  we have for the radius and velocity at $t_{\rm{sf}}$ 

\begin{equation}
{R_{{\rm{sf}}}} = 13.38\ {\left( {\frac{{{E_{51}}}}{{\mu \,{n_0}}}} \right)^{0.2}}t_{{\rm {sf4}}}^{\,0.4}\,\,\,{\rm{(pc)}} 
\label{eq4}
\end{equation}
and

\begin{equation}
\upsilon _{\rm sf} = 5.24 \times {10^7}\,\,{\left( {\frac{{{E_{51}}}}{{\mu \,{n_0}}}} \right)^{ 0.2}}{t_{\rm sf4}^{ - 0.6}\,{\rm (cm/s)}},
\label{eq5}
\end{equation}
 where ${t_{{\rm{sf4}}}}$ is  ${t_{{\rm{sf}}}}$ in ${10^4}$yr, $\mu $ is the mean ISM molecular weight in units of the proton mass. At the moment ${t_{{\rm{sf}}}}$ the formation of cool thin shell ends, and the following expansion of this shell takes place due to the pressure of the hot interior gas. Because we are interested mainly in the SNRs in Sedov and radiative phases,  we consider here the evolution of these shells in detail, taking into account the pressure of the ISM. Unlike the adiabatic case, where the expansion rate of the gas immediately behind the shock front lags behind the shock velocity, for a cold isothermal shell these speeds are the same.  The system of equations governing the expansion of this snow-plowing shell includes the equation of mass conservation \citep{Bis95}, 

\begin{equation}
\frac{{dM}}{{dt}} = 4\pi \,{\rho _0}R_{\rm{s}}^2\,{\upsilon _{\rm{s}}},
\label{eq6}
\end{equation}
the equation of momentum conservation,

\begin{equation}
\frac{{d(M \cdot {\upsilon _{\rm{s}}})}}{{dt}} = 4\pi R_{\rm{s}}^2({P_{{\rm{in}}}} - {P_0}),
\label{eq7}
\end{equation}	
and the equation of energy conservation, 

\begin{equation}
\frac{{d{P_{{\rm{in}}}}}}{{dt}} =  - 3\,\gamma {P_{{\rm{in}}}} \cdot \frac{{{\upsilon _{\rm{s}}}}}{{{R_{\rm{s}}}}},
\label{eq8}
\end{equation}
where $M = (4\pi /3) \cdot {\rho _0}R_{\rm{s}}^3$ is the mass accumulated in the thin shell, $\,{\upsilon _{\rm{s}}} = d{R_{\rm{s}}}/dt$ is the velocity of the shell, which is equal to the shock velocity; ${P_{in}}$ and ${P_0}$ are the total pressures inside the shell and in the ISM. 
With a little manipulation the system of Eqs. (\ref {eq6})  and (\ref {eq7}) can be transformed to an equation linear in ${\upsilon_{\rm{s}} ^2}$, 
\begin{equation}
\frac{{d{\upsilon_{\rm{s}} ^{\rm{2}}}}}{{dR}} + \frac{{6{\upsilon_{\rm{s}} ^{\rm{2}}}}}{R} = \frac{{6{P_0}}}{{{\rho _0}}} \cdot \frac{1}{R}\left( {\frac{{{P_{in}}}}{{{P_0}}} - 1} \right).
\label{eq10}
\end{equation}
	
By using Eq.~(\ref{eqMN}), the general solution of the Eq.~(\ref{eq10}) can be written as

\begin{equation}
\gamma M_s^2 = \frac{2}{{2 - \gamma }}\frac{{{P_{1in}}}}{{{P_0}}} \cdot \frac{{R_1^{3\gamma }}}{{R_s^{3\gamma }}} + \frac{C}{{R_s^6}} - 1,
\label{eq11}
\end{equation}
where $C$ is the integration constant, ${P_{\rm 1in}}$ is the pressure inside of the shell at arbitrary radius ${R_{\rm s1}}$ of the radiative remnant. For this radius we take the radius $R_{\rm sf}$.  
According to the exact Sedov solution, the ratio of the pressure immediately behind the shock front  to the central almost uniform pressure is constant,  and for $\gamma  = 5/3$ this ratio is $\alpha  \approx 0.31$ \citep{s77}. 
   Therefore, for the initial pressure inside the remnant, which in the course of evolution will change according to Eq.~(\ref{eq8}) and push the dense shell forward, we have  

\begin{displaymath}
{P_{1in}} = \alpha \, \cdot \frac{2}{{\gamma  + 1}}{\rho _0}\upsilon _{{\rm{sf}}}^{\rm{2}}{\rm{(}}{R_{{\rm{sf}}}}{\rm{)}} = \frac{{2 \, \alpha  \, \gamma }}{{\gamma  + 1}}{P_0}M_{{\rm Rad}}^2, 
\end{displaymath}
where 
$ M_{\rm Rad}^{} = {\upsilon _{{\rm  sf}}}/{c_{{\rm ms0}}} $ 
is the shock Mach number at the time ${t_{{\rm{sf}}}}$. 
Numerical calculations show  \citep{cioffi88,MS74}  that  at the transition between Sedov and radiative phases a sudden $\sim 20\% $ decrease in shock velocity occurs.   Therefore the initial velocity of the shell at the moment ${t_{{\rm{sf}}}}$, when the snow-plow phase just begins to act, is   ${\upsilon _1} = \theta  \cdot {\upsilon _{{\rm{sf}}}}$, with $\theta  \approx 0.8$ \citep{DrW91}. 

After substituting these last expressions into Eq.~(\ref{eq11}), we have 

\begin{equation}
C = \left\{ {\left[ {{\theta ^2} - \frac{{4 \cdot \alpha }}{{(2 - \gamma )(\gamma  + 1)}}} \right] \, \gamma  \, M_{{\rm{Rad}}}^2 + 1} \right\} \cdot R_{{\rm{sf}}}^6.
\label{eq12}
\end{equation}

Finally, we  rewrite the Eq.~(11) as the dependence of the shock Mach number on the normalized shock radius $x \equiv R_{\rm s}/{R_{{\rm{sf}}}}$  in the form

\begin{equation}
{M_{\rm s}} = \frac{1}{{\sqrt \gamma  }}\sqrt {\frac{A}{{{x^{3\gamma }}}} - \frac{B}{{{x^6}}} - 1},	
\label{eq13}
\end{equation}

where  
\begin{displaymath}
A = \frac{{4\,\alpha \,\,\gamma }}{{(2 - \gamma )(\gamma  + 1)}} \cdot {M}_{{\rm{Rad}}}^2,
\end{displaymath}
and

\begin{displaymath}    
B = A - \gamma \,{\theta ^2}\,{M}_{{\rm{Rad}}}^{\rm{2}} - 1.
\end{displaymath}

Eq.~(\ref{eq13}) can be used to describe the evolution of SNR. The dependences ${M_{\rm s}} {(R_{\rm s})} $ and ${\upsilon_{\rm s}} {(R_{\rm s})}$ can be numerically converted into time dependences with the help of integrals:

\begin{equation}
t - {t_{{\rm{sf}}}} = \int\limits_{{R_{{\rm{sf}}}}}^{R_{\rm s}} {\frac{{dR'}}{{{\rm{v}}(R')}} = \frac{{{R_{{\rm{sf}}}}}}{{{c_{\rm ms0}}}}\,\,\int\limits_1^{R_{\rm s}/{R_{{\rm{sf}}}}} {\frac{{dx}}{{{M_{\rm s}}(x)}}} }.
\label{eq14}
\end{equation}

\begin{figure}
\centering
\includegraphics[width=9.5cm]{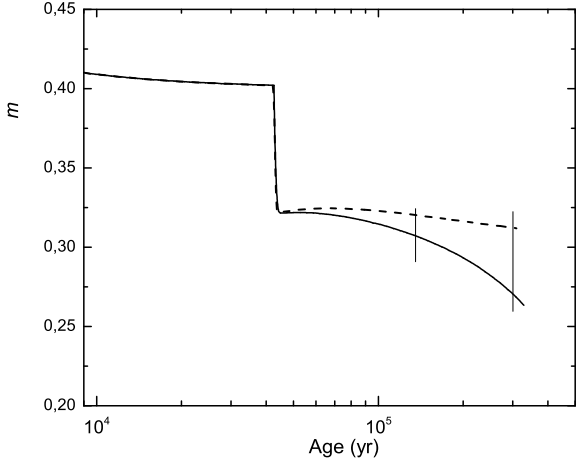}

\caption{Dependence of the expansion parameter $m$ on time for an SNR with  $E_0 = 10^{51}{{\rm erg}},\,\,\, {n_0} = 0.5\,\,{\rm cm}^{ - 3}$ and $P_{04}=4\times {10^4}~\,{\rm{K}}\, {\rm{c}}{{\rm{m}}^{ - 3}}$. The solid line corresponds to the case of evolution with the effect of pressure of the ISM, the dashed line for the case without pressure. The vertical lines denote the moments when the shock Mach number $M_{\rm s} = 2$ (left) and $M_{\rm s} = 1$(right).}
\label{fig1}
\end{figure}

In principle, Eq.~( \ref{eq13}) can be analyzed analytically, but since there are no physically meaningful asymptotes, we used it in numerical calculations. Assuming a power-law dependence for the expansion law as ${R_{\rm s}} \propto {t^m},$ the dependence of 
\begin{displaymath}    
m = \frac{{{M_{\rm s}} \cdot {c_{\rm ms0}}t}}{{{R_{\rm s}}}} 
\end{displaymath} 
on time is presented in Fig.~\ref{fig1}. This parameter is called the expansion parameter and it measures the shell deceleration.

The figure shows that $m\sim 0.4$ in the Sedov phase, at the end of which it suddenly drops to a slowly varying value of $0.3-0.33$ during the remaining 
active life of the SNR ($M_{\rm s}>2$) in accordance with detailed numerical results \citep[e.g][]{ch1974}. However, our solution does not show the value $m = 2/7 \approx 0.29$ following from the pressure-driven snowplow solutions \citep{mo77,Blin82}. 
The late-time behavior of the SNRs is mainly determined by the conditions in the ISM, where the remnant evolves.  
For a more detailed analysis, it is desirable to determine the lifetime of the remnant by using the shock Mach number. This parameter shows how strong the shock wave is, which in turn determines the visibility of the remnant in any waveband of the electromagnetic spectrum. We define the lifetime of the remnant as the moment when the shock Mach number becomes equal to $2$. Indeed, an SNR with $M _{\rm s} < 2$ is unlikely to be detected in any wavelength range, and such SNRs can be considered as dead SNRs. When we take  the effect of the interstellar pressure into account, the occurrence of the so-called momentum-conserving snowplow with ${R_s} \propto {t^{1/4}}$, known as the Oort phase, becomes problematical.  
 The late-time evolution problem was studied  in detail by Cioffi et al.~(1988), who also discussed whether an evolution  of the remnant without the radiative phase is at all possible. In our case this situation occurs when ${M_{{\rm{rad}}}} \le 2$, which can be expressed as  

\begin{displaymath}
{M_{{\rm{rad}}}} = 30.6 \, {\mu ^{3/10}} \, {\gamma ^{ - 1/2}} \, E_{51}^{1/14} \, n_0^{9/14} \,P_{04}^{ - 1/2},
\end{displaymath}
from which the condition on the density can be obtained (for $\gamma  = 5/3;\,\,\mu  \approx 0.61)$
\begin{displaymath}
{n_{\rm{0}}} \le 0.03 \, E_{51}^{ - 1/9} \, P_{04}^{7/9}.
\end{displaymath}
In these equations $P_{04}$ is the total pressure in the ambient ISM in units of $10^4~\,{\rm{K}}\, {\rm{c}}{{\rm{m}}^{ - 3}}$.
As we noted, at these densities 
the evolution of the shock radius can be described by the Sedov law of expansion $R_{s} \sim t^{2/5}$ up to values of the Mach number of $ M_{\rm s}\sim 2 $, although at these values of $M_{\rm s}$ the internal structure strongly deviates from the self-similar exact solution \citep{s77,CA82}.

It is important to note that in the evolution of old and evolved SNRs the external pressure of the ISM becomes a dominant factor that determines the lifetime and maximum achievable size of the SNR. In general, the total pressure is the sum of contributions of thermal particles, cosmic rays, and magnetic fields at various scales.
It is well known that the thermal pressure in disk galaxies is only a small part of the total pressure in the gas; in particular, \citet{bcox} have shown that 
the turbulent pressure is considerably higher than the thermal pressure.They have estimated the total pressure (with thermal plus turbulent contributions) of the interstellar medium to be on the order of $10^4~\,{\rm{K}}\, {\rm{c}}{{\rm{m}}^{ - 3}}$. For the subsequent analysis, it is important to note that it follows from the general theoretical analysis that the spatial variation of the total pressure in the plane of the galaxy is about an order of magnitude or less \citep[e.g.][]{wol03}, while the contrast in other physical characteristics of the ISM (density, temperature) can reach 4 - 5 orders of magnitude. Therefore, it is reasonable to assume that the total pressure of the ISM is in the range of $(1-5)\times {10^4}~\,{\rm{K}}\, {\rm{c}}{{\rm{m}}^{ - 3}}$.

\begin{figure}
\centering
\includegraphics[width=9.5cm] {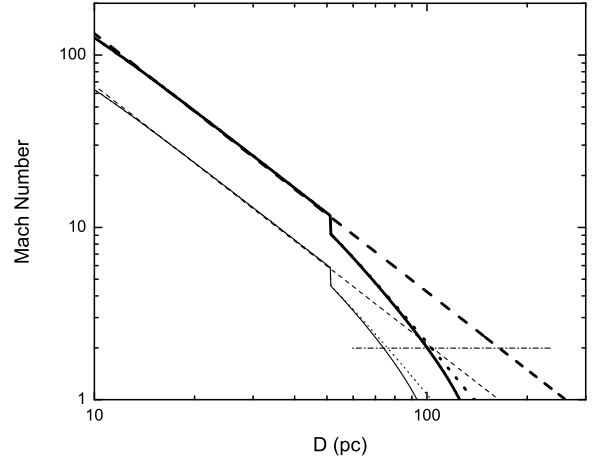}
\caption{Dependence of the Mach number on the diameter of the SNR for ${E_0} = {10^{51}}{\rm{erg}}$ and ${n_0} = 0.5\,\,{\rm{c}}{{\rm{m}}^{ - 3}}$ for the two values of pressure ${P_{04}} = 2.17$ (thick lines) and ${P_{04}} = 8.6$ (thin lines). The solid and dotted curves describe the solution of the system of Eqs. (6-8) with and without the pressure of the ISM, and the dashed curves describe the Sedov solution without radiative cooling. The horizontal dash-dotted line corresponds to the point ${M_{\rm{s}}} = 2$.}
\label{fig2}
\end{figure}

In Fig.~\ref{fig2} some curves describing the evolution of the shock Mach number with the diameter are shown, which demonstrate the role of the interstellar pressure on the evolution of the SNR. The pressure also affects the evolution of SNR through the definition of the shock Mach number, the parameter that serves as the measure of the blast wave intensity.  The pressure effect becomes noticeable at the very end of the SNR life and, in general, the including the interstellar pressure results in a considerable reduction of the highest values of the SNR diameter and lifetime.

 As we have  proposed  above, the active life of  the SNR ends when its Mach number reaches the value of 2. We adopted the diameter and age of the remnant at this moment as the largest size $D_{\rm  max }$ and longest lifetime of the SNR. Obviously, this choice is arbitrary, but at this value of $M_{\rm s}$, in specific conditions of the ISM, the SNR can  preserve its integrity more or less, but its ability to generate detectable emission in any of the energy bands is strongly reduced.  
As can be seen in Fig.~\ref{fig2} the value of largest diameter of the SNR in real conditions of the ISM is much larger than the observational values of 30 - 50 pc. 
This fact implies that either we do not see a large portion of real existing objects, or that  in real conditions of the ISM SNRs do not expand to large diameters, that is, our model is incorrect.

Using this solution for an SNR evolving in the ISM with ${n_0} = 1\,{\rm{c}}{{\rm{m}}^{ - 3}}$  and  $P_0 = 2.2 \times {10^4}\,\,{\rm{K}} \, {\rm{c}}{{\rm{m}}^{ - 3}}$,  we   estimate the largest diameter and longest lifetime  to be  90 pc and $4.5 \times {10^5}$ yr, respectively. These values are reduced to 68 pc and $1.75 \times {10^5}$ yr, respectively, when the pressure is  increased to ${P_0} = 8.6 \times {10^4}\,{\rm{K}}\, {\rm{c}}{{\rm{m}}^{ - 3}}$.  If we take for the density ${n_0} = 0.1\,{\rm{c}}{{\rm{m}}^{ - 3}}$, which is typical for the warm phase of the ISM,  the largest diameter and longest lifetime of the SNR are 126 pc and  $2.11 \times {10^5}$ yr  at   $P_0 = 2.2 \times {10^4}\,\,{\rm{K}} \, {\rm{c}}{{\rm{m}}^{ - 3}}$ and 102 pc and $1.11 \times {10^5}$ yr when the interstellar pressure is $P_0 = 8.6 \times {10^4}\,{\rm{K}}\, {\rm{c}}{{\rm{m}}^{ - 3}}$. From these numerical estimations we can also see that the effect of the ambient interstellar pressure on the active lifetime of the SNR is more prominent than on the value of its largest size.

\section {X-ray SNRs}

Supernova remnants are studied practically in all bands of the electromagnetic emission, but the most extensive observational information is available in radio, X-ray, and optical wavebands. 
For the present study the main advantage of the X-ray band over others is that the nature of X-ray emission of evolved SNRs is clearly established, which we cannot say about the nature of the emission in other wavebands. The thermal nature of the SNR X-ray emission is often used as one of the main properties of shell-type SNRs - to identify the X-ray sources as SNRs, they should be extended objects and sources of thermal emission. Indeed, the majority of detected shell-like SNRs exhibits thermal X-ray emission. The exception is the small number of young SNRs with power-law X-ray spectra that are generally believed to originate from synchrotron emission, although bremsstrahlung emission of electrons with a non-Maxwellian distribution function 
\citep{Asvarov1990}
cannot be excluded.  

Supernova remnants are relatively well studied in X-ray wavebands, though the details of this emission are somewhat complicated because the X-ray spectrum in the 0.1-2 keV range is formed by a variety of line and continuum processes that depend on the details of the temperature and density structure and on the physical conditions in the hot gas, and on the interstellar absorption as well. Another difficulty is the inhomogeneity of data from different data sets   
of X-ray SNRs. 
In this sense, for statistical studies of SNRs the radio observations are favorable, as is the case with our Galaxy and the Magellanic Clouds. Unfortunately, because we lack a reliable theory of the origin of the  SNR radio emission, we cannot simulate their evolution in the present study. For the same reason we cannot use the rich data of optical observations of SNRs that are located mostly in neighboring galaxies.

\subsection {Data}

With regard to the above discussion, the recent work of \citet{L10} (L10), devoted to a statistical study of SNRs in optical and X-ray ranges in the nearby galaxy M33, is of great value for the present study. The SNRs in M33, as is typically the case for nearby galaxies, are mainly detected in the optical waveband. The previously published list of SNRs \citep{Gordon1998, Gordon1999}, which consisted of 99 optical SNRs, has been expanded  to 137 SNRs and SNR candidates in L10. 

Using Chandra data from the ChASeM33 survey, 82 of 137 SNR candidates were detected as X-ray sources. According to the authors, the catalog includes all the SNRs in the portions of M33 covered by the ChASeM33 survey with 0.35-2 keV X-ray luminosity higher than $\sim 4 \times {10^{34}}\rm\, { erg\,\, s^{-1}}$, and the sample in L10 provides the largest sample of remnants detected at optical and X-ray wavelengths in any galaxy, including the Milky Way.
 For uniformity of the statistics it is desirable to have results of relatively simultaneous observations conducted with the help of the same instrument. Fortunately, the catalog of SNRs in M33 from L10 satisfies these conditions. Importantly,  the X-ray data in this catalog were obtained with the help of the Chandra observatory between 2005 September and 2006 November (L10).
  We also used the recent ChASeM33 point-source catalog of M33
 \citep[T11 hereafter]{T11}, 
in which only 45 SNRs of the 82 SNRs from L10 are detected as  point sources. These SNRs are expected to be less contaminated with background emission, which makes this sample of SNRs much more statistically uniform.

	We note that the ChASeM33 survey only covers the inner region of M33 to a radius of about 4.3 kpc, which is about $70\%$ of the area of the galaxy \citep{pluc08}. 


\subsection {X-ray emission from SNRs}

For a Maxwellian energy distribution of the electrons, which we assume is the case for the SNRs discussed here, the emissivity of the thermal continuum X-ray emission at photon energy ${E_x}$  is given \citep{vink12} (in units of   ${\rm{erg}}\,{{\rm{s}}^{ - 1}}{\rm{c}}{{\rm{m}}^{ - 3}}{\rm{H}}{{\rm{z}}^{ - 1}}$)

\begin{equation}
 {\varepsilon _{ff}} = \frac{{{2^5}\pi {e^6}}}{{3{m_e}{c^3}}}{\left( {\frac{{2\pi }}{{3k{m_e}}}} \right)^{1/2}}{g_{ff}}\,\,{n_e}\sum\limits_i {{n_i}Z_i^2}  \cdot T_e^{ - 1/2}\exp \left( { - \frac{{{E_x}}}{{k{T_e}}}} \right),
\label{eq15}
\end{equation}
where with ${g_{ff}} \approx 1$, the gaunt-factor, which depends weakly on the electron temperature ${T_e}$ and photon energy ${E_x}$, ${n_e}$ and ${n_i}$ are the densities of electrons and ions with charge Z, respectively, and the remaining variables have standard meanings.  Due to the factor  ~$\exp \left( { - {E_x}/k{T_e}} \right)$ in this formula, the X-ray flux from the SNR is very sensitive to the temperature of the post-shock plasma and the range of photon energies at which the X-ray detector  operates. 
The temperature of the post-shock electrons is determined by the evolutionary status of the remnant's shock wave, namely, by the shock velocity. If we assume a strong shock in a fully ionized gas with complete temperature equilibration between ions and electrons, then the X-ray temperature implies a shock velocity \citep{mckeehol}

\begin{displaymath}
 {\upsilon _{\rm{s}}} = (\gamma  + 1)\,\,{\left( {\frac{{k{T_{\rm{s}}}}}{{2(\gamma  - 1){m_{\rm{p}}}{\mu _{\rm{s}}}}}} \right)^{1/2}} = 715\,\,{\left( {{T_{{\rm{s}}{\rm{,keV}}}}/{\mu _{\rm{s}}}} \right)^{1/2}}\,\,{\rm{km}}\,{{\rm{s}}^{ - 1}},
\end{displaymath}
 where ${T_{{\rm{s}}{\rm{,keV}}}}$ is the post-shock temperature in keV and ${\mu _s}$ is the post-shock mean mass per free particle in proton masses ${m_{\rm{p}}}$. 
 If the temperature of the plasma falls below the lower energy limit of the X-ray detector (in the present analysis this is $ 0.35$ keV), the flux of X-ray photons will decrease exponentially as the radius of the remnant increases and the remnant disappears as an X-ray source.  So, the SNRs in our list are the remnants with ${\upsilon _{\rm{s}}} \ge 715{\left( {0.35/{\mu _{\rm{s}}}} \right)^{1/2}}\,\,{\rm{km}}\,{{\rm{s}}^{ - 1}} \approx 530\,{\rm{km}}\,{{\rm{s}}^{ - 1}}$. 
We did not consider the complicated behavior of the electron temperature behind the shock front \citep[see e.g.,][]{perviz13} 
or the role of radiation in the lines at low energies of the observed band.
 We assumed that SNRs emit in X-rays according to Eq.~(\ref{eq15}), although at energies $\sim 0.1$ keV the contribution of the line emission to the total emission may be very substantial. 
\begin{figure}
\includegraphics[width=9.0 cm] {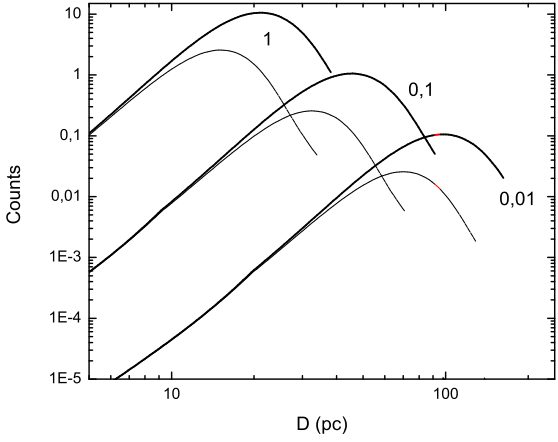} 
\caption{Evolution of the $[0.35 - 1.1]$ keV (thick curves) and $[1.1 - 2.6]$ keV (thin curves) X-ray photon count rates as a function of SNR diameter for the ambient particle densities $1$, $0.1$, and $0.01$   ${\rm{c}}{{\rm{m}}^{ - 3}}$, as denoted near the corresponding curves, and ${E_0} = {10^{51}}$ erg,  ${N_H} = {10^{21}}\,{\rm{c}}{{\rm{m}}^{ - 2}}$ as in the case of M33. 
}
\label{fig31}
\end{figure}

The X-ray detector counts the photons within specific instrument energy channels. To convert these data into physically important information (photon flux, energy spectrum, luminosity, etc.), we have to perform a number of intermediate steps and make assumptions about the spectra of the electrons that emit in the X-ray waveband. All these make the obtained  information less reliable. Therefore, we used the raw count rates given in the list of SNRs in L10 and made no assumptions about the spectra of the sources other than the assumption that the emissivity has a thermal nature  
 of the form (Eq.~\ref{eq15}). 
We calculated one of the important characteristics of the X-ray emission of the remnant - the count rate of X-ray photons in the range of energies $[{E_{x1}};{E_{x2}}]$,  as 
\begin{eqnarray}
\Phi ({E_{x1}};{E_{x2}}) = A_{1} \cdot \int\limits_V {dV} \int\limits_{{E_{x1}}}^{{E_{x2}}} {d{E_x}\,} {n_e}\sum\limits_i {{n_i}Z_i^2}  \cdot T_e^{ - 1/2}E_x^{ - 1}
\nonumber
\\
\times\exp \left( { - \frac{{{E_x}}}{{k{T_e}}} - \sigma ({E_x}) \cdot {N_H}} \right).
\label{eq16}
\end{eqnarray}

To calculate this quantity,   we used the approximation $\sigma ({E_x}) = \left( {{c_0} + {c_1}{E_x} + {c_2}E_x^2} \right)E_x^{ - 3} \times {10^{ - 24}}\,\,{\rm{c}}{{\rm{m}}^2}$ for the cross-section per hydrogen atom from \citet{zombeck}, where ${E_x}$ is given in keV and the coefficients ${c_0},\,{c_1}$ and  ${c_2}$  are tabulated. For the radial dependences of the densities and the temperature in this formula we have used the exact Sedov self-similar solution when the shock wave was strong (${M_s} \ge 10$), but for ${M_{\rm s}} < 10$ 
we used the approximate solution of Cox \& Anderson (1982). The evolution  of the X-ray count rates with the SNR diameter is shown in Fig.~\ref{fig31}.     This figure also illustrates how the observed values of the  X-ray emission depend on the characteristics of the energy band at which the emission is detected. The analysis of  Eq.~(\ref{eq16}) shows that the  highest X-ray count rate of the  SNR is reached 
when the temperature of the X-ray emitting electrons corresponds to the energy band of the observations.
Applying this picture to the SNRs in M33, we used for the distance and column density the values of 817 kpc and ${N_H} = {10^{21}}\,{\rm{c}}{{\rm{m}}^{ - 2}}$  (L10), respectively.   

 To reduce the effect of the poorly determined normalizing coefficient $A_{1}$ in Eq.~(\ref{eq16}), it is useful to employ the hardness ratio (HR), which is commonly calculated as the normalized difference of the exposure-corrected counts in two energy bands. The dependence of the HR on the energy of photons directly indicates the nature of the X-ray emission - thermal, power-law, or some other.  
In L10 this parameter was defined as the ratio 
$ HR = {[\Phi (1.2;2.6) - \Phi (0.35;1.2)}]/{{\Phi (0.35;8.0)}}$
where $\Phi (0.35;1.2),\,\,\Phi (1.2;2.6)$, and $\Phi (0.35;8.0)$ are the numbers of counts correspondingly in the $0.35-1.2$ keV, $1.2-2.6$ keV and $0.35-8.0$ keV bands, and for all the sources detected in X-rays the values of $HR$ hardness ratios are given. To select the SNRs from other Chandra sources, L10 used the criterion that $HR < 0$.  This directly reflects the thermal nature of the X-ray emission of detected SNRs. The evolution of $HR$ with diameter is determined mainly by the behavior of the exponential factor $\sim\exp \left( { - {E_x}/k{T_e}} \right)$, therefore the distribution of this parameter can be used to estimate the evolutionary status of the objects in the list.  

\begin{figure}
\centering
\includegraphics[width=9.5cm]{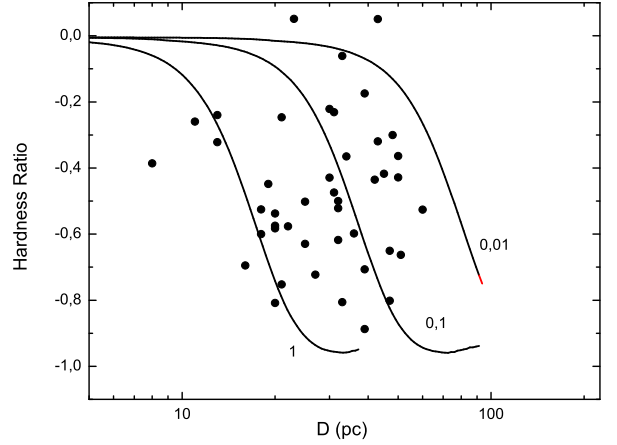}
\caption{Hardness ratio (HR, see text for definition) evolution  with diameter of SNR for  ${E_0} = {10^{51}}~{\rm{erg}}$ and ${n_0} = 0.01,\,\,0.1,\,\,{\rm{and}}\,\,\,1.0\,\,{\rm{c}}{{\rm{m}}^{ - 3}}$. Filled circles correspond to the 44 SNRs from the  T11 catalog.
}
\label{fig41}
\end{figure}

\begin{figure}
\centering
\includegraphics[width=8.5cm]{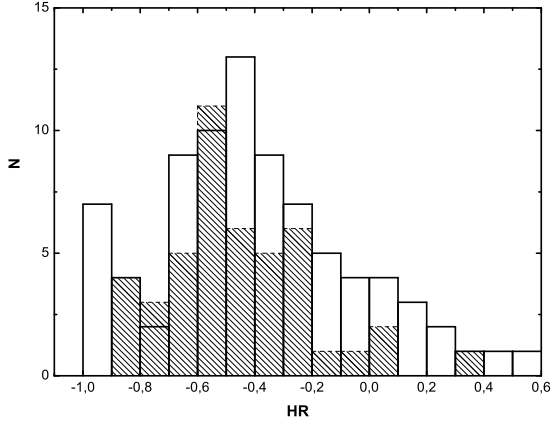} %
\caption{Hardness ratio distribution of SNRs in M33. The solid line histogram corresponds to 82 SNRs from L10, the histogram for 45 SNRs with HRs calculated by using the data from T11 is drawn with dashed lines.}
\label{fig51}
\end{figure}

We also calculated the HRs using the recently published catalog of X-ray sources in M33 T11, where the count rates in different energy bands are given. 
The hardness ratios calculated from these data are  well correlated with the HRs given in L10, with the exception of the source L10-096 (G98-73) (source number in the catalog T11 is 426), for which we obtained $HR=0.34$, but L10 found $HR=-0.6$. The only SNR with a high positive value of $HR \sim 0.4$ in both catalogs is the object L10-119 (Scr. No in T11 is 493).  In Fig.~\ref{fig41} the evolution of $HR$ with $D$ is shown together with 44 observational points calculated using the data from T11; for the SNR L10-096 we adopted $HR=-0.6$ as in L10, and L10-119 is not shown in this figure. The hardness ratio histogram is shown in Fig.~\ref{fig51}, where HRs for 45 SNRs 
from T11 and HRs of 82 SNR from L10 are presented. In both cases it can be seen that there is a higher concentration of SNRs around $HR\sim -0.5$. The mean values of the HR are $-0.44$ for both calculations. It is important to note that at $HR \sim -0.5$ the SNR reach the highest X-ray luminosity, which is why we assume that the population of observed SNRs with $HR$ around $-0.5$ forms the uniform set of objects that are unaffected by selection effects.

\section{Model: results and discussion}

Observations show that SNRs have very different characteristics. This is partly due to the scatter in the properties of parent SNe (mainly, SN energies in case of evolved sources), partly   due to the broad spread of the ISM properties (density, pressure, chemical abundances, etc.) where  SNRs expand. 

To separate the effects of these factors on the statistics of SNRs, we built a simple evolution model of SNRs  with different initial birth parameters and evolving in sites with different ambient conditions. We generated a set of SNR as follows: every $\Delta T$ year an SN occurs, the initial parameters of which are randomly distributed  in intervals: 1) the explosion energy (kinetic) $E_{0}=(0.2 \div 5) \times {10^{51}}$ erg; 2) the ejecta mass $M_{\rm {ej}}=(1 \div 5)$ $M_{\sun}$. The parameters of the ISM, the total pressure and the  density,  where the SN occurs, also were chosen in a random manner: 3) the total pressure in the range $P_{0}=(1\div 5)\times {10^4} \,\rm {K\,cm^{-3}}$; 4) for the density we considered a three-phase model in which for the hot, warm and cold  phases the values of the density are taken randomly from the intervals $n_{h}=[0.005\div 0.1)\, \rm{cm^{-3}}$, $n_{w}=[0.1\div 1)\, \rm {cm^{-3}}$,  and $n_{c}=[1\div 10]\, \rm{cm^{-3}}$ with volume-filling factors $\phi_{h}$, $\phi_{w}$, and $\phi_{c}$, respectively. For the cold phase we adopted the constant value $ \phi_{c}=0.01$, but $\phi_{h}$ or $\phi_{w}$  ($\phi_{h}+\phi_{w}=0.99$) was the main varying input parameter of the model.      

In this way, both populations of SNRs, the observed and the modeled, are sets of SNRs with different birth characteristics, and these SNRs are evolving in different conditions of the ISM and are at a different stage of evolution, but there is one main difference between them - the modeled SNRs are free from any selection effects.

We assumed that the shock of radius $R_s$  expands in accordance with Eqs. (\ref{eq2}) and  (\ref{eq13}) into a fully ionized  gas with a ratio of specific heats $\gamma  = 5/3 $ and with a helium abundance relative to hydrogen as $n_{\rm He}= 0.1\,n_{\rm H}$.  The number density  and electron density were $n=n_{\rm He}+n_{\rm H}=1.1\,n_{\rm H}$  and  $n_{\rm e}= n_{\rm H} + 2\,n_{\rm He}=1.2\,n_{\rm H}$, respectively. 

\begin{figure}
\centering
\includegraphics[width=9.0 cm] {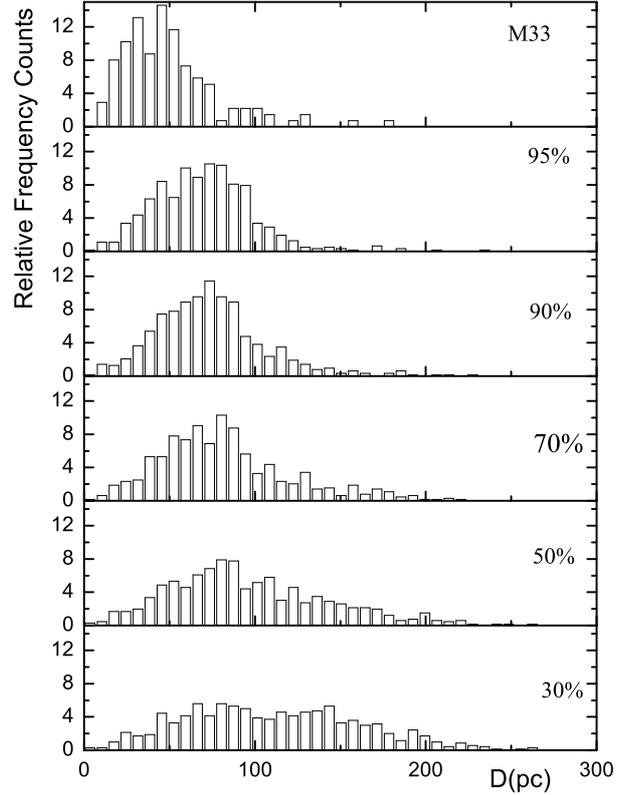}%
\caption{Size distribution of observed SNRs in M33 and Monte Carlo simulated SNRs for various values of the warm phase filling-factor (indicated in the panels). In simulations $\Delta T=250 $ yr, the lifetime is determined as $t = t(M_s = 2)$ and the largest dimension of simulated SNRs is $D = D(M_{\rm s} = 2)$. The bin size is 7~pc.}
\label{fig61}
\end{figure}

As an example, we show in Fig.~\ref{fig61} the size distributions of SNRs modeled for different values of the filling factor of the warm phase of the ISM. This figure illustrates the general behavior of the size distribution of SNRs: with the decrease of the warm-phase filling factor the size distribution broadens and the fraction of the large diameter SNRs increase.    
The distribution of 137 real SNRs in M33 is also shown in Fig.~\ref{fig61}. However, from this comparison of modeled and observed distributions we are unable to obtain useful information either on the SNR evolution or about the ISM in M33. This is because the sample of simulated SNRs is free from selection effects but the experimental SNRs are the objects detected in a certain electromagnetic waveband and suffer from various selection effects. In general, a selection effect occurs when we cannot see a weak point source or when we overlook a weak, extended source because of background emission.
Typically, an object bright in some waveband may be dim in other waveband, therefore observations at different wavebands are of great importance for statistical consideration. Accordingly, to compare the modeled size distribution with the observed distribution in a given waveband we need to model the evolution of the SNR emissivity in this waveband.

As we showed, SNRs with $HR\sim -0.5$ constitute a more or less uniform group of SNRs, therefore  we selected SNRs with these values of $HR$ among the observed as well as the modeled SNRs. In Table 4 of L10, of the 82 X-ray SNRs and SNR candidates the HR lies in the interval $ - 0.7 \le HR \le  - 0.3$ for 48 objects. This subset of SNRs have diameters ranging from 8 to 111 pc; the mean diameter is 40 pc, the median diameter is 36 pc. The statistics of these SNRs is given in Table~\ref{snrs}. In this table  ${D_{\rm m}}$, $D_{\rm min }$, and $D_{\rm max}$ are the mean, minimum, and maximum diameters of SNRs in each sample, respectively.   In the
last column we list the ratio of highest to lowest values of the ($0.35-2.6$) keV  count rates, which were determined as the ratio of the count rates averaged over the three brightest SNRs to the count rates averaged over the three weakest sources in the list of each sets of SNRs.

\begin{table}
\caption{The parameters of observed and modeled SNRs, selected according to the criterion of $ - 0.7 \le HR \le  - 0.3$. }
\label{snrs}
\medskip
\centering 
\begin{minipage}{11.5cm}
\begin{tabular}{lrrrrrr} \hline

\hline
 & ${D_{\rm m}}$   &	$D_{\rm min }$   &	$D_{\rm max }$      &   $\frac {Count_{\rm   max}}{Count_{\rm  min}}$\\  

\hline
&  	(pc) 		& 	(pc)  	& 	(pc) 	&  		&  	\\
\hline
Observed SNRs 	&			&		&		&		&	\\
All 137 SNRs	&	49.6$\pm$28.8	&		&		&		&	\\
82 X-ray SNRs	 & 	40.3$\pm$20.6 	& 	8 	&	111		&	261\\ 
48 SNRs   		& 	39.9$\pm$22.6	&	8	&	111	&	197\\
 	&			&		&		&		&	\\

Modeled SNRs 	&			&		&		&		&	\\
Run1		&	41.0$\pm$20.6	&	16	&	109		&	179\\  
Run2		&	39.2$\pm$21.4	&	16.1	&	124	&	88\\  
Run3		&	39.0$\pm$19.5	&	12.1	&	104		&	380\\  
Avrg over 20 runs	&	39.1$\pm$18.4	&	13.1	&	107		&	210\\  

\hline
\end{tabular}
\end{minipage}
\end{table}

\begin{figure}
\centering 
\includegraphics[width=8.5cm]{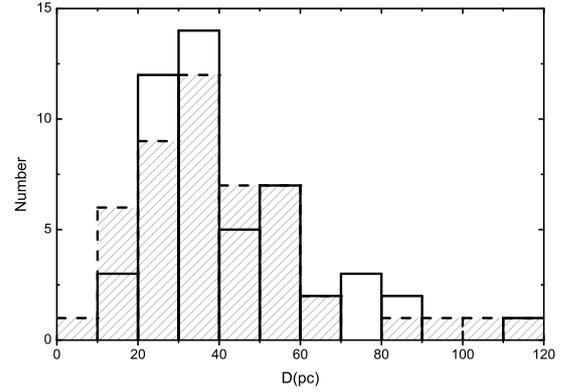} %
\caption{ Diameter distribution of SNRs with $ - 0.7 \le HR \le  - 0.3$ in M33 (dashed lines) and modeled SNRs (solid lines). The data for 48 SNRs in M33 are taken from L10, the 49 modeled SNRs represent the scenario of $90\%$ warm gas filling-factor and a birthrate of 150 yr. The size of the bins is 10 pc }
\label{fig7}
\end{figure}

Returning to our model, for each supernova remnant along with kinematic parameters such as the velocity, radius, Mach number, etc., we calculated the ($0.35 - 2.6$)~keV photon counts with the help of Eq.~(\ref{eq16}) and thermal X-ray luminosity in the same energy range with arbitrary normalization.  
Initially, for the birthrate, $(\Delta{T})^{-1}$, we took higher values to reduce the statistical fluctuations, and the generation of new SNRs stops when the birth and death rates become equal. As discussed above, the moment when the shock Mach number drops to 2 was adopted as the SNR death point. 
The main input parameter of each run was the filling-factor of the hot (or warm) phase of the ISM. In the generated set of SNRs we selected the subset of SNRs with HR in $ - 0.7 \le HR \le  - 0.3$, which then we compared with the selected subset of observed SNRs. As initial fitting parameters we used the mean value of the diameters and the ratios of highest to lowest count rates and X-ray luminosities. To take the influence of statistical fluctuations into account, which are inherent in the Monte Carlo method, the same model was repeated on the fit procedure with different statistical realizations of the model with the same input parameters.  Results of several runs with  $\phi_{w}=0.90$, differing only by the value of the random seed, 
are presented in Table~\ref{snrs}, and in Fig.~\ref{fig7}  the diameter distribution histogram is shown for comparative purposes. 

The observed count rates for detected SNRs vary by a factor of 430, which we used as the parameter in fitting the modeled and observed SNRs. In the modeled SNR sample the ratio $\frac {Count_{\rm   max}}{Count_{\rm  min}}$
is very close to the observed value (Tabl.~\ref{snrs}). This indicates that the observed SNR sample with HR $\sim-0.5$ constitutes a statistically complete set of objects, that is, there are no missing  low-luminosity SNRs.
The best agreement between the simulated and observed distributions was reached at a filling-factor of the warm phase of $90\%$. From the comparison of the number of  observed and modeled SNRs with HR in $ - 0.7 \le HR \le  - 0.3$ we obtain the value $\Delta{T} \sim (140-150)$ yr for SNRs in M33. This value agrees well with the estimate of 
\citet{Tamman94} for a birthrate  of 147 years between SN events in M33 and  is consistent with the  fact that for the past 100 years of observations  no  SN event has been detected.

\begin{figure}
\centering 
\includegraphics[width=8.5cm]{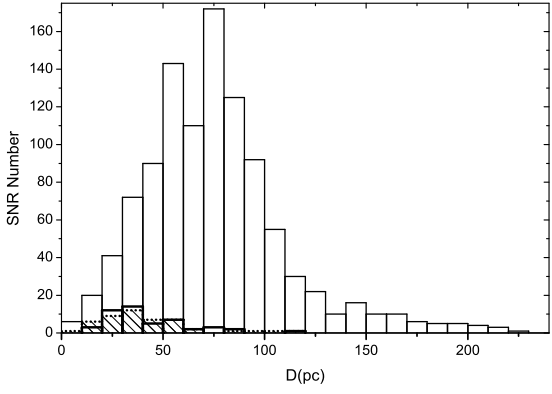} %
\caption{ Diameter distribution of 1050 modeled SNRs with $ M_{s}\geq 2$ together with the distributions shown in Fig.~\ref{fig7}. For the modeled SNRs the filling-factor of the warm phase of the ISM is  $90\%$, the birthrate is 1/150 yr$^{-1}$, and  ${P_0} = 8.6 \times {10^4}\,{\rm{K}}\, {\rm{c}}{{\rm{m}}^{ - 3}}$. }
\label{fig8}
\end{figure}

 Knowing the birthrate of SNRs allows us to estimate their total number, the volume of the galaxy occupied by them, and other very important characteristics of the galaxy. To estimate the full number of SNRs in M33 it is important to know the lifetime of the SNR in dependence on the initial and environmental conditions. 
As was discussed above, if we assume that the end of the SNR life occurs when the shock Mach number reaches $2$, then the total number of SNRs will be dependent on the total pressure in the ISM. In M33 the total number of SNRs with $M_{s}\geq 2$ lies in the interval $(1000 - 1600)$ SNRs when we adopt a pressure in the range  $(2-8)\times {10^4}~\,{\rm{K}}\, {\rm{c}}{{\rm{m}}^{ - 3}}$.  In Fig.~\ref{fig8} the distribution of all modeled SNRs with $M_{s}\geq 2$ for $8\times {10^4}~\,{\rm{K}}\, {\rm{c}}{{\rm{m}}^{ - 3}}$ is shown together with the distributions of observed and modeled SNRs selected according to the criteria  $ - 0.7 \le HR \le  - 0.3$ shown in Fig.~\ref{fig7}. 
Here we have taken the highest attainable ambient pressure in the ISM of M33 to obtain the minimum possible number of active SNRs.  
It is important to emphasize that neither the birth rate of SNRs nor the volume-filling factor is dependent on the ambient pressure. At higher ambient interstellar  pressure the lifetime of the SNR decreases significantly, which results in a decrease of the number of active SNRs. At the same time, the increase of the pressure results in an insignificant decrease of the maximum size of the remnant (see Sec.2). Therefore, the effect of the interstellar ambient pressure on the size distribution is negligibly small.

As can be seen in Fig.~\ref{fig8}, the diameter distribution of the SNRs that are predicted by the model but that are undetected peaks at $70 - 80$ pc,
which is almost twice the mean diameter of observed X-ray active SNRs.  They cover up to $5\%$ of the surface of the Galaxy, and assuming the volume of the observed part of M33 is $V=4\cdot\pi\cdot R^{2}\times h=4\cdot\pi\cdot4.3^{2}\times1\,\,({\rm{kpc}}^{3})=6.86\times10^{66}\,\,{\rm{cm}^{3}}$,  SNRs with $M_{s}\geq 2$ occupy less than $1\%$ of the total volume of the galaxy.

\begin{figure}
\centering
\includegraphics[width=8.5cm]{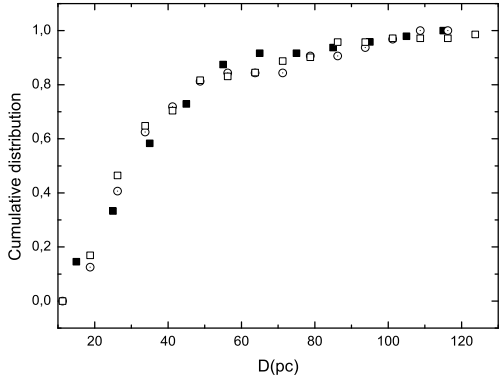} %
\caption{Normalized cumulative $N-D$ distribution for X-ray SNRs with $ - 0.7 \le HR \le  - 0.3$  in M33 (filled squares) and two representative (run 1-open squares, run 3-open diamonds) modeled SNR subsets. The data for 48 SNRs in M33 are taken from L10; for the modeled SNRs the filling-factor of the warm phase of the ISM is  $90\%$ }
\label{fig9}
\end{figure}

  With the aim of comparing of the modeled and observed statistics of SNRs with $ - 0.7 \le HR \le  - 0.3$, we show in Fig.~\ref{fig9} their normalized cumulative number $ - $ diameter ($N - D$) distributions. As can be seen from this comparison, the correspondence between observed and modeled distributions is very good. For remnants with $D < 60\,$pc the cumulative $N - D$ relation for both SNR subsets (observed and modeled) is approximated by the form $N(D)\sim {D^\eta }$, with $\eta  = 1.2 - 1.6$, though for the Sedov expansion in the uniform ISM the expected value is $\eta  = 2.5$. This long-known discrepancy  \citep{Mathewson1983,Mathewson1984} is now explained  by the strong variations in ambient density \citep{Berk1987} and partly by the selection effects \citep{Hug84}, but our analysis shows that 
	the selection effects play a minor role.
	
	\begin{figure}
\centering
\includegraphics[width=8.5cm]{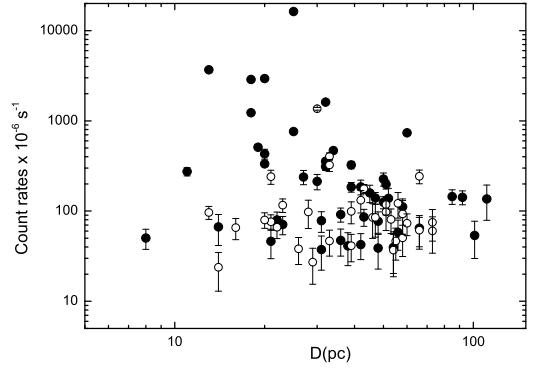} %
\caption{ Dependence of $0.35 - 2 \,\,{\rm {keV}}$ photon count rates of M33 SNRs on the diameters. The 48 SNRs with $ - 0.7 \le HR \le  - 0.3$ are denoted with filled circles, open circles denote the remaining 34 SNRs. All the data are taken from the list in L10.}
\label{fig10}
\end{figure}

\begin{figure}
\centering
\includegraphics[width=8.5cm]{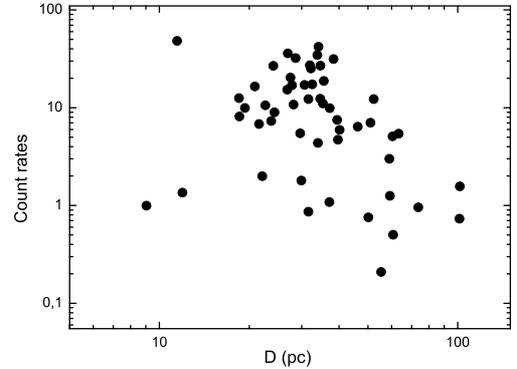} %
\caption{ Dependence of $(0.35 - 2) \,\,{\rm {keV}}$ photon count rates on the diameter for modeled SNRs. The normalization is arbitrary. In this run for the $90\%$ warm-phase filling-factor and the mean total pressure $4\times {10^4}~\,{\rm{K}}\, {\rm{c}}{{\rm{m}}^{ - 3}}$ the number of SNRs with $M_{\rm{s}}\geq 2$ is $1530$, of which 53 SNRs with $ - 0.7 \le HR \le  - 0.3$ are presented in this figure.}
\label{fig11}
\end{figure}

	We considered the dependence of SNR X-ray luminosity on diameter.  
	Fig.~\ref{fig10} displays the relation between the count rates (in place of X-ray luminosity) and the SNR diameters. 
	Obviously, there is practically no correlation between these parameters. Almost the same relationship holds for the simulated SNRs (see  Fig.~\ref{fig11}).
	The lack of correlation between count rates and the diameters can be easily explained by the fact that the observed set of SNRs consists of objects evolving in very different initial and environmental conditions, although every individual SNR has a count rate - diameter relation as shown in Fig.~\ref{fig31}.   
It is important to note  that the subset of considered remnants constitutes a  homogeneous statistical set of objects free from selection effects, which means that the lack of  an $L_x -D$ dependence is the common picture for the shell-like SNRs, and can be used as a strong evidence in favor of the considered picture of SNR evolution in the real conditions of the ISM. 

The modeled SNR sample, which agrees very well with the observed X-ray SNRs, also exhibits  the well-known correlation of apparent density  $n_0$ with SNR diameter.  This correlation, to the author's knowledge,  was first noticed by \citet{mo77} for a  small sample of galactic SNRs. Although they used this correlation to support a multiphase model of the ISM, we now know that this relationship has a statistical nature \citep{Berk1987}. The statistical relation $D -n_0$ for the modeled sample has the form $D\sim {n_{0} }^{-0.33}$ with the correlation coefficient  $\sim 0.77$, which does not contradict the result of \citet{Berk1986} of  $D\sim {n_{0} }^{-0.39\pm 0.04}$.  

\section{ Conclusions}

We have modeled the statistics of the SNRs evolving in the three-phase ISM.  The evolution of SNRs in various conditions of the ISM was considered, taking into account the effect of the total interstellar pressure. We assumed that the X-ray emission from the SNRs is thermal. Using the Monte Carlo method, we then modeled the set of SNRs by varying the values of the initial (explosion energy and ejected mass of the SN) and environmental parameters (conditions in the ISM), which control the evolution and final fate of the SNR. Among many other statistical relations, the size distribution of SNRs is very sensitive to the density distribution in the ISM. In the three-phase ISM the shape of this relation strongly depends on the filling factor of phases of the ISM. 

This simple method is only effective for a statistically complete and homogeneous sample of observed SNRs. 
Fortunately, the recently published detailed catalog of SNRs
in the nearby galaxy M33 by \citet{L10}~(L10) contains all the information needed for our analysis. 

We showed that the population of SNRs with an X-ray hardness ratio of about $- 0.5$ yields the statistically most complete sample of objects that is free from various selection effects. We selected 48 such SNRs from the total numbers of 82 X-ray  SNRs in the list of L10 for a comparison with the modeled SNRs with the same HR values. From the comparison of these two sets of SNRs we obtained the following  results:	\begin{itemize}
		\item the volume-filling factor of the warm phase of the ISM of M33 is $\sim90\%$; 
		\item the birthrate of SNRs in M33 is  6.7 $- $ 7.1 SNRs per millennium for the inner $4.3$~kpc. 
	\end{itemize} The estimate of the total number of SNRs in M33 depends on the total interstellar pressure of the galaxy and, also, on the definition of the SNR lifetime. The SNR was considered to exist when the shock Mach number was equal to or higher than 2. Our model predicts the existence of at least 1000 such alive SNRs in M33, which occupy less than $1\% $ of the galaxy volume within $R<4.3$~kpc.

The modeled set of supernova remnants shows several statistical relations that also exist in the  
observed populations of SNRs, such as the shape of the relation  $N -  D$, the  negative correlation between the apparent ambient density $n_{0}$  and diameter $D$,  and the lack of a statistical correlation between  X-ray luminosity  and diameter.

\begin{acknowledgements} 

This work was supported by the Science Development Foundation under the President of the Republic of Azerbaijan –  Grant No EIF-2010-1(1)-40/05. The author is grateful to the referee, Elly Berkhuijsen, for her careful reading and invaluable comments and suggestions, which greatly improved this paper.
\end{acknowledgements}


\end{document}